\documentclass[conference]{IEEEtran}
\IEEEoverridecommandlockouts
\input{befe}

\usepackage{algorithmic}
\usepackage{graphicx}
\pgfplotsset{compat=1.18}
\usepgfplotslibrary{statistics}
\usetikzlibrary{patterns}
\usetikzlibrary{positioning}
\usepackage{siunitx}
\usepackage{tabularx}

\newacronym{mimo}{MIMO}{multiple-input multiple-output}
\newacronym{simo}{SIMO}{single-input multiple-output}
\newacronym{mse}{MSE}{mean square error}
\newacronym{cme}{CME}{conditional mean estimator}
\newacronym{pdf}{PDF}{probability density function}
\newacronym{adc}{ADC}{analog-to-digital converter}
\newacronym{mmse_abstract}{MMSE}{minimum mean square error}
\newacronym{mmse}{MMSE}{minimum mean square error}
\newacronym{snr}{SNR}{signal-to-noise ratio}
\newacronym{evd}{EVD}{eigenvalue decomposition}
\newacronym{crb}{CRB}{Cram\'er-Rao bound}
\newacronym{map}{MAP}{maximum a posteriori}
\newacronym{cs}{CS}{compressive sensing}
\newacronym{ls}{LS}{least squares}
\newacronym{awgn}{AWGN}{additive white Gaussian noise}
\newacronym{csi}{CSI}{channel state information}
\newacronym{ml}{ML}{maximum likelihood}
\newacronym{pmf}{PMF}{probability mass function}
\newacronym{cdf}{CDF}{cumulative distribution function}
\newacronym{rv}{RV}{random variable}
\newacronym{gmm}{GMM}{Gaussian mixture model}
\newacronym{mt}{MT}{mobile terminal}
\newacronym{bs}{BS}{base station}
\newacronym{acg}{AGC}{automatic gain control}
\newacronym{ula}{ULA}{uniform linear array}
\newacronym{upa}{UPA}{uniform planar array}
\newacronym{mfa}{MFA}{mixture of factor analyzer}
\newacronym{vae}{VAE}{variational autoencoder}
\newacronym{em}{EM}{expectation-maximization}
\newacronym{dft}{DFT}{discrete Fourier transform}
\newacronym{elbo}{ELBO}{evidence lower bound}
\newacronym{mmwave}{mmWave}{millimeter wave}
\newacronym{rf}{RF}{radio frequency}
\newacronym{gamp}{GAMP}{generalized approximate message passing}
\newacronym{iht}{IHT}{iterative hard thresholding}
\newacronym{fft}{FFT}{fast Fourier transform}
\newacronym{psd}{PSD}{positive semi-definite}
\newacronym{nn}{NN}{neural network}
\newacronym{relu}{ReLU}{rectified linear unit}
\newacronym{uatf}{UF}{use-and-then-forget}
\newacronym{mrc}{MRC}{maximum-ratio combining}
\newacronym{los}{LOS}{line-of-sight}
\newacronym{nlos}{NLOS}{non-line-of-sight}
\newacronym{3gpp}{3GPP}{3rd Generation Partnership Project}
\newacronym{cnn}{CNN}{convolutional NN}
\newacronym{aqnm}{AQNM}{additive quantization noise model}
\newacronym{vi}{VI}{variational inference}
\newacronym{vamp}{VAMP}{vector AMP}
\newacronym{gan}{GAN}{generative adversarial network}
\newacronym{dm}{DM}{diffusion model}
\newacronym{flop}{FLOP}{floating point operation}

\def\BibTeX{{\rm B\kern-.05em{\sc i\kern-.025em b}\kern-.08em
    T\kern-.1667em\lower.7ex\hbox{E}\kern-.125emX}}

\newcommand{\lineWidth}{1.2pt}
\newcommand{\marksize}{2.2pt}
\pgfplotsset{tick label style={font=\small},label style={font=\small},legend style={font=\scriptsize}}

\pgfplotsset{
  every axis legend/.append style={
    font=\scriptsize 
  }
}

\pgfplotsset{
  every axis title/.append style={
    font=\scriptsize 
  }
}

\pgfplotsset{every axis/.append style={
                    label style={font=\scriptsize},
                    tick label style={font=\scriptsize}  
                    }}

\definecolor{myblack}{RGB}{70,70,70}
\definecolor{myblue}{RGB}{65,105,225}
\definecolor{mygreen}{RGB}{0,139,139}
\definecolor{myorange}{RGB}{255,150,0}
\definecolor{myred}{RGB}{255,69,0}
\definecolor{mylila}{RGB}{153,50,204}

\glsdisablehyper

\usetikzlibrary{external}
\begin{document}
\bstctlcite{IEEEexample:BSTcontrol}

\title{Enhancing Channel Estimation in Quantized Systems with a Generative Prior
\thanks{The authors acknowledge the financial support by the Federal Ministry of
Education and Research of Germany in the program of ``Souver\"an. Digital.
Vernetzt.''. Joint project 6G-life, project identification number: 16KISK002.}
}
\author{
	\centerline{Benedikt Fesl, Aziz Banna, and Wolfgang Utschick}\\
	\IEEEauthorblockA{School of Computation, Information and Technology, Technical University of Munich, Germany\\
	Email: \{benedikt.fesl, aziz.banna, utschick\}@tum.de
    }
}

\maketitle

\begin{abstract}
    Channel estimation in quantized systems is challenging, particularly in low-resolution systems. In this work, we propose to leverage a \ac{gmm} as generative prior, capturing the channel distribution of the propagation environment, to enhance a classical estimation technique based on the \ac{em} algorithm for one-bit quantization. Thereby, a \ac{map} estimate of the most responsible mixture component is inferred for a quantized received signal, which is subsequently utilized in the \ac{em} algorithm as side information. Numerical results demonstrate the significant performance improvement of our proposed approach over both a simplistic Gaussian prior and current state-of-the-art channel estimators. Furthermore, the proposed estimation framework exhibits adaptability to higher resolution systems and alternative generative priors.
\end{abstract}


\begin{figure}[b]
\onecolumn
\centering
\copyright \scriptsize{This work has been submitted to the IEEE for possible publication. Copyright may be transferred without notice, after which this version may no longer be accessible.}
\vspace{-1.3cm}
\twocolumn
\end{figure}
\section{Introduction}
Utilizing a generative prior for inverse problems in wireless communications has attracted considerable attention recently due to the high potential of improved estimation performance, e.g., using \acp{gmm} \cite{koller22gmm}, \acp{mfa}~\cite{fesl23mfa}, \acp{vae}~\cite{baur2024leveraging}, \acp{gan}~\cite{9252921}, or \acp{dm}~\cite{MIMO_channel_estimation_diff,fesl2024diffusionbased}.
These works promise a significant performance improvement over classical techniques and learning-based regression \acp{nn} by learning the channel distribution for a whole radio propagation environment, capturing valuable prior information. 
A general limitation is the common assumption of infinite precision \acp{adc} at the receiver. 
The primary motivation for deploying low-resolution \acp{adc} is the significant improvement in energy efficiency, particularly crucial in massive \ac{mimo} systems.
However, utilizing a generative prior for low-resolution quantization at the receiver, introducing a challenging inverse problem due to the pronounced nonlinearity introduced by the low-resolution \acp{adc}, is not very well studied. 

An exception is the work in \cite{fesl24quant}, where conditionally Gaussian latent generative models are utilized to parameterize a conditional Bussgang estimator. Although this work offers promising performance gains for channel estimation, it is limited by the linearity of the parameterized Bussgang estimator, based on the observation that the Bussgang estimator deteriorates from the \ac{mse}-optimal \ac{cme}, especially for higher dimensions \cite{fesl23cme}.

A different approach was discussed in \cite{mezghani_em,stoeckle_em}, focusing on the \ac{map} optimization problem instead of the \ac{mmse} estimator, which is solved with an \ac{em} algorithm. It was shown that explicit knowledge of the prior distribution is essential, and the case of a simple Gaussian prior is discussed, leading to closed-form solutions. 
However, a simplistic Gaussian prior considerably underrepresents the channel distribution of a whole propagation environment.

\textbf{Contributions:}
We propose a novel estimation framework for quantized systems, enhancing the \ac{em} algorithm from \cite{mezghani_em,stoeckle_em} by utilizing a conditionally Gaussian latent model as generative prior for learning the channel distribution of a whole radio propagation environment, providing valuable information for the subsequent estimation task. Due to the conditional Gaussianity of the leveraged model, we get an analytic closed-form solution for the corresponding M-step of the \ac{em} algorithm, resulting in reduced computational complexity. 
While our study concentrates on a \ac{gmm} as generative prior and the extreme case of one-bit quantization, the proposed estimation framework exhibits adaptability for higher resolution systems and alternative generative priors, e.g., \acp{mfa} or \acp{vae}. We discuss different structural constraints on the covariances of the \ac{gmm}, i.e., Toeplitz or circulant structure, resulting in reduced memory overhead and computational complexity.
Numerical results demonstrate significant performance gains compared to using a simplistic Gaussian model as prior and state-of-the-art channel estimators in quantized systems.

\textbf{Notation:} 
The $i$-th element of a vector $\B x$ is denoted as $[\B x]_i$ and its real and imaginary parts are given as $\B x_\Re$ and $\B x_\Im$, respectively. The standard Gaussian \ac{cdf} is denoted by $\op \Phi(x) = \int_{-\infty}^x \tfrac{1}{\sqrt{2\pi}}\exp(-\tfrac{x^2}{2})\op d x$.

\section{System and Channel Model}
Consider an uplink transmission of $P$ pilot signals from single-antenna \acp{mt} to an $N$-antenna \ac{bs}, operating one-bit \acp{adc}. The quantized receive signal is
$\B R = Q(\B Y) = Q(\B h \B a\T + \B N)$,
where $\B R = [\B r_1,\dots, \B r_P]\in \mathbb{C}^{N \times P}$, $\B Y$ is the unquantized receive signal, $\B h\in \mathbb{C}^{N}$ denotes the wireless channel, and $\B a\in\mathbb{C}^P$ is the pilot vector with the entries, cf. \cite{fesl24quant}, 
$[\tilde{\B a}]_i = \beta_i \exp\left(\op j \frac{\pi}{2P}(i-1)\right),~i\in\{1,\dots,P\}$,
where $\beta_i = \frac{1}{2} + \frac{i-1}{2(P-1)}$ is the amplitude spacing. The pilot vector is subsequently normalized to fulfill the power constraint $\|\B a\|_2^2=P$.
Furthermore, $\B N= [\B n_1,\dots, \B n_P]\in \mathbb{C}^{N\times P}$ is white Gaussian noise with $\B n_i\sim \NC(\B 0, \sigma^2\eye)$ and $Q$ denotes the complex-valued one-bit quantization function
\begin{align}
    Q(\B y) = \frac{1}{\sqrt{2}}\left(\sign(\B y_\Re) + \op j \sign(\B y_\Im)\right).
    \label{eq:1bit_quant}
\end{align}
After column-wise vectorization, the system model is
\begin{align}
    \B r = Q(\B y) = Q(\B A\B h + \B n) \in \C^{N P}
    \label{eq:sytem_model}
\end{align}
with $\B r  = \vect(\B R)$, $\B y  = \vect(\B Y)$, $\B n  = \vect(\B N)$, and $\B A = \B a \otimes \eye$. 
By normalizing the channels as $\op E[\|\B h\|_2^2] = N$, the \ac{snr} is defined as $\text{SNR} = 1/\sigma^2$.

We utilize a stochastic-geometric channel model based on the \ac{3gpp} spatial channel model~\cite{3gpp,NeWiUt18}, where channels are modeled conditionally Gaussian: $\B h| \B \delta \sim \NC(\B 0, \B C_{\B h | \B\delta})$.
The random vector $\B \delta$ collects the uniform distributed angles of arrival/departure and path gains of the main propagation clusters between a \ac{mt} and the \ac{bs}.
The \ac{bs} employs a \ac{ula} such that the spatial channel covariance matrix is given by
\begin{equation}\label{eq:3gpp_cov}
    \B C_{\B h| \B \delta} = \int_{-\pi}^\pi \omega(\gamma; \B \delta) \B t(\gamma) \B t(\gamma)\h \op d \gamma
\end{equation}
with $\B t(\gamma) = [1, \op e^{\op j\pi\sin(\gamma)}, \dots, \op e^{\op j\pi(N-1)\sin(\gamma)}]\T$
being the array steering vector for an angle of arrival $\gamma$, and $\omega$ is a power density consisting of a sum of weighted Laplace densities whose standard deviations describe the angle spread of the propagation clusters~\cite{3gpp}.
To create a training dataset $\mathcal{H}$, we generate random angles and path gains for every data sample $\B h_t$, combined in $\B \delta_t$, and then draw the sample as $ \B h_t \sim \NC(\B 0, \B C_{\B h | \B \delta_t}) $, which results in an overall non-Gaussian channel distribution \cite{9842343} of the training dataset $ \mathcal{H} = \{\B h_t\}_{t=1}^{T_{\text{train}}}.$

\section{EM Algorithm with a Gaussian Prior}\label{sec:em}
In this section, we briefly summarize the \ac{em} algorithm for channel estimation from \cite{mezghani_em,stoeckle_em}, where the case of a Gaussian prior is explicitly discussed.
The goal is to solve the \ac{map} optimization problem, given as
\begin{align}
    \hhat_{\text{MAP}} = \argmax_{\B h\in \C^N} p(\B h | \B r) = \argmax_{\B h\in \C^N} p(\B h, \B r).
    \label{eq:map}
\end{align}
However, as outlined in \cite{mezghani_em,stoeckle_em}, the direct optimization of \eqref{eq:map} is intractable since the involved densities are inaccessible even under a known (Gaussian) prior. 
Thus, an \ac{em} algorithm is utilized, where the unquantized receive signal $\B y$ is considered as a latent variable. The corresponding joint likelihood $p(\B r, \B h, \B y)$ is decomposed as $p(\B r, \B h, \B y) = p(\B r| \B y, \B h) p(\B y| \B h) p(\B h)$
where $p(\B r | \B y)$ is an indicator function since the quantization function~\eqref{eq:1bit_quant} is deterministic and $p(\B y | \B h) = \NC(\B y; \B A \B h, \sigma^2\eye)$. In this section, it is further assumed that the prior distribution is a zero-mean Gaussian, i.e., $p(\B h) = \NC(\B h; \B 0, \B C_{\B h})$.

After some reformulations, the E-step of the \ac{em} algorithm in the $\ell$-th iteration is shown to consist of the computation of the conditional expectation $\hat{\B y}^{(\ell)} = \E[\B y | \B r, \B h^{(\ell)}]$, whose real part can be computed elementwise as, cf. \cite{mezghani_em,stoeckle_em},
\begin{align}
\begin{aligned}
\label{eq:e-step}
    [\hat{\B y}^{(\ell)}_\Re]_i = [\B b^{(\ell)}_\Re]_i + \frac{\sigma}{\sqrt{2\pi}} 
    \frac{[\B r_\Re]_i \exp\left(-\frac{1}{\sigma^2}[\B b_\Re^{(\ell)}]_i^2\right)}{\op \Phi\left(\frac{2}{\sigma}[\B r_\Re]_i [\B b^{(\ell)}_\Re]_i\right)} 
    \end{aligned}
\end{align}
where $\B b^{(\ell)} = \B A \B h^{(\ell)}$. The imaginary part is computed analogously.
Although the Gaussian \ac{cdf} has no analytic expression, there exist practicably feasible approximations, e.g.,~\cite{1094433}.

Furthermore, in the M-step, the updated channel estimate $\hhat^{(\ell + 1)}$ is computed as \cite{stoeckle_em}
\begin{align}
    \hhat^{(\ell + 1)} = \argmin_{\B h \in \C^N} \|\B A \B h - \hat{\B y}^{(\ell)}\|_2^2 - \sigma^2 \log p(\B h).
    \label{eq:m_step}
\end{align}
In the case of a zero-mean Gaussian prior, as discussed above, the optimization problem \eqref{eq:m_step} has the closed-form analytic solution~\cite{mezghani_em,stoeckle_em}
\begin{align}
    \label{eq:m_step_sol}
    \hhat^{(\ell + 1)} = \left(\B A\h \B A + \sigma^2 \B C_{\B h}\inv\right)\inv \B A\h \hat{\B y}^{(\ell)}.
\end{align}
The E- and M-steps are repeated after properly initializing $\hhat^{(0)}$ until convergence.

\section{Enhanced Channel Estimation with a Generative Prior}\label{sec:generative_prior}

The prerequisite of the \ac{em} algorithm in \Cref{sec:em} that the channel is Gaussian imposes a severe limitation. This becomes particularly evident when considering the channel distribution of a whole radio propagation environment, generally being considerably underrepresented by a simplistic Gaussian prior. Thus, we aim to enhance the presented \ac{em} algorithm by learning the channel distribution with a generative model, providing a much stronger prior.
A variety of generative models exist for implicitly learning the channel distribution through a \ac{nn} that produces valid channel samples, e.g., \acp{gan} \cite{9252921} or \acp{dm} \cite{MIMO_channel_estimation_diff,fesl2024diffusionbased}; however, the M-step in \eqref{eq:m_step} may no longer be tractable to solve, rendering such models difficult to use. 

To circumvent this problem, we resort to conditionally Gaussian latent variable models, i.e., \acp{gmm}, \acp{mfa}, or \acp{vae}. These models facilitate the inference of a latent variable that, when conditioned on the channel distribution, produces an analytically tractable conditionally Gaussian model for the channel. In this study, our emphasis is primarily on \acp{gmm}, although the other models are, in principle, equally suitable.

A \ac{gmm} is a \ac{pdf} of the form
\begin{align}
\label{eq:gmm}
    p(\B h) = \sum_{k=1}^K p(k) \NC(\B h; \B \mu_{\B h|k}, \B C_{\B h | k})
\end{align}
with the set of learnable parameters $\{p(k), \B \mu_{\B h|k}, \B C_{\B h|k}\}_{k=1}^K$ being the mixing coefficients, means, and covariances of the corresponding mixture components.
The parameters of the \ac{gmm} are fitted in an offline phase via the \ac{em} algorithm\footnote{Note that this \ac{em} algorithm has no intended connection to the presented algorithm in \Cref{sec:em}.} for a given training dataset $\mathcal{H}$ of channel samples, cf. \cite[Ch. 9]{bookBi06}.
An essential property of \acp{gmm} is that for a given data sample, the \textit{responsibility} of each component can be computed as, cf. \cite[Ch. 9]{bookBi06}, $p(k | \B h) \propto p(k) \NC(\B h; \B \mu_{\B h| k}, \B C_{\B h|k})$.
The \ac{gmm} can be described via a \textit{discrete} latent variable with a categorical distribution, which conditions on one of the $K$ components \cite[Ch. 9]{bookBi06} and, thus, yields a conditionally Gaussian latent variable model. In \cite{fesl24quant}, it was shown that a zero-mean \ac{gmm} with $\B \mu_{\B h| k} = \B 0$ for all $k=1,\dots,K$ is sufficient to well approximate a feasible wireless channel distribution. Since this allows for simplified expressions of the resulting algorithm, we also use a zero-mean \ac{gmm}.

Directly plugging the learned \ac{gmm} distribution into \eqref{eq:e-step} yields no closed-form solution, requiring, e.g., a gradient descent technique. Solving this non-convex optimization problem in every iteration leads to high complexity and the possibility of reaching local minima.
Thus, we instead propose to infer the most responsible \ac{gmm} component for a given pilot observation and run the \ac{em} algorithm from \Cref{sec:em} for the conditional Gaussian prior, leading to a closed-form analytic solution. An estimate of the \ac{gmm} components' responsibility for a given pilot observation $\B r$ from \eqref{eq:sytem_model} can be computed as \cite{fesl24quant}
\begin{align}\label{eq:pk_given_r}
    \hat{p}(k | \B r) = \frac{p(k) \NC(\B r; \B 0, \B C_{\B r|k})}{\sum_{i=1}^K p(i) \NC(\B r; \B 0, \B C_{\B r|i})}
\end{align}
where $\B C_{\B r|k}$ is computed through $\B C_{\B y|k} = \B A \B C_{\B h|k}\B A + \sigma^2 \eye$ with the arcsine law \cite{272490} as 
\begin{equation}\label{eq:arcsin_law}
	\begin{aligned}
		\B C_{\B r| k} = \frac{2}{\pi}\left( \arcsin\left(\Re(\B R_{\B y|k})\right) 
		+ \op j \arcsin\left(\Im(\B R_{\B y|k})\right)\right)
	\end{aligned}
\end{equation}
where $\B R_{\B y|k} = \diag(\B C_{\B y|k})^{-\frac{1}{2}}\B C_{\B y|k} \diag(\B C_{\B y|k})^{-\frac{1}{2}}$.

Consequently, the most responsible \ac{gmm} component for a given quantized pilot observation is the \ac{map} estimate \cite{10318056}
\begin{align}
    k^\star = \argmax_k \,\hat{p}(k| \B r).
\end{align}
When assuming the corresponding most responsible \ac{gmm} component is the true prior distribution, the M-step in the $\ell$-th iteration of the \ac{em} algorithm from \Cref{sec:em} changes to 
\begin{align}
    \hhat^{(\ell + 1)} = \argmin_{\B h \in \C^N} \|\B A \B h - \hat{\B y}^{(\ell)}\|_2^2 - \sigma^2 \log p(\B h | k^\star).
    \label{eq:m_step_gmm}
\end{align}
Due to the Gaussianity of $p(\B h|k^\star) = \NC(\B 0, \B C_{\B h | k^\star})$ by design, the closed form solution is given as
\begin{align}
    \hhat^{(\ell + 1)} = \left(\B A\h \B A + \sigma^2 \B C_{\B h|k^\star}\inv\right)\inv \B A\h \hat{\B y}^{(\ell)}.
\end{align}
The E-step in \eqref{eq:e-step} is unchanged since it does not depend on the prior distribution.
The offline phase, consisting of fitting the \ac{gmm} once to the underlying channel distribution of the radio propagation environment and the online evaluation of the \ac{em} algorithm, enhanced by the generative prior, is summarized in Algorithm~\ref{alg:em_enhanced}, labeled \textbf{GMM-EM}.

\begin{algorithm}[t]
\caption{Training and inference of the GMM-EM estimator.}
\label{alg:em_enhanced}
\begin{algorithmic}[1]
\renewcommand{\algorithmicensure}{\textbf{Offline Training Phase}}
\ENSURE
\REQUIRE $\mathcal{H} = \{\B h_t\}_{t=1}^{T_{\text{train}}}$, $K$
\STATE Train zero-mean GMM: $\{p(k), \B C_{\B h|k}\}_{k=1}^K$
\par\vskip.5\baselineskip\hrule height .4pt\par\vskip.5\baselineskip
\renewcommand{\algorithmicensure}{\textbf{Online Channel Estimation Phase}}
\ENSURE
\REQUIRE $\{p(k), \B C_{\B h|k}\}_{k=1}^K$, $\B r$, $\sigma^2$, $\B A$, $L_{\text{max}}$, $\gamma$
\STATE Compute responsibilities $\{\hat{p}(k | \B r)\}_{k=1}^K$ via \eqref{eq:pk_given_r}
\STATE Compute \ac{map} estimate: $k^\star = \argmax_{k} \hat{p}(k| \B r)$
\STATE Initialize with \ac{ls} estimate: $\hhat^{(0)} = \B A^\dagger \B r$
\FOR{$\ell = 0$ {\bfseries to} $L_{\text{max}}$}
    \STATE \textbf{E-step:} Compute $\hat{\B y}^{(\ell)} = \E[\B y | \B r, \hhat^{(\ell)}]$ via \eqref{eq:e-step}
    \STATE \textbf{M-step:} $\hhat^{(\ell+1)} = (\B A\h \B A + \sigma^2\B C_{\B h | k^\star}\inv )\inv \B A\h \hat{\B y}^{(\ell)}$
    \IF{$\|\hhat^{(\ell+1)} - \hhat^{(\ell)}\|_2 \leq \gamma \|\hhat^{(\ell)}\|_2$}
    \STATE \textbf{break}
    \ENDIF
\ENDFOR
\end{algorithmic}
\end{algorithm}

\subsection{Structural Constraints}
To further incorporate structural properties to the generative prior, imposed by the antenna structure at the \ac{bs}, i.e., Toeplitz (\textbf{GMM-EM toep}) or circulant (\textbf{GMM-EM circ}) covariances, we utilize a structured \ac{gmm} as outlined in \cite{fesl22gmm}. Thereby, each component's covariance is described by the decomposition
\begin{align}
    \B C_{\B h | k} = \B Q\h \diag(\B c_{\B h|k}) \B Q,
\end{align}
where $\B Q$ is a (oversampled) \ac{dft} matrix and $\B c_{\B h|k} \in \R^N_+$. These structural constraints allow for having fewer model parameters and a lower online complexity overhead of the resulting estimator due to the usage of \acp{fft}.

\subsection{Memory and Complexity Analysis}
The number of model parameters is scaling with $\mathcal{O}(KN^2)$ or $\mathcal{O}(KN)$ for the \ac{gmm} with full or structurally constrained covariances, respectively. 
For reducing the computational complexity, many iterative computations in Algorithm~\ref{alg:em_enhanced} can be precomputed since the \ac{gmm} is fixed after the offline phase, e.g., the filter in the M-step and the inverses $\B C_{\B r|k}\inv$ for evaluating the responsibilities \eqref{eq:pk_given_r}. Thus, the order of complexity is dominated by matrix-vector products, yielding $\mathcal{O}(\max(L,K)PN^2)$, where $L$ is the number of iterations of the \ac{em} algorithm until convergence. We note that the computations of the $K$ responsibilities \eqref{eq:pk_given_r} are trivially parallelizable. Although the order of complexity is the same for the structurally constrained versions, the number of \acp{flop} is drastically reduced by using \acp{fft}.

\section{Baseline Channel Estimators}
\label{sec:baselines}
In the following, we introduce several theoretical and practical state-of-the-art baselines that are most related and relevant for evaluating the estimation performance of the proposed algorithm:
\begin{itemize}
    \item \textbf{genie-EM:} 
    Assuming genie-knowledge of the covariance matrix $\B C_{\B h|\B \delta}$ from \eqref{eq:3gpp_cov} for each pilot observation allows for running the \ac{em} algorithm with the ground-truth prior by replacing $\B C_{\B h}$ with $\B C_{\B h|\B \delta}$ in the M-step \eqref{eq:m_step_sol}. This yields a theoretical bound on the best-possible estimation performance when utilizing the \ac{em} approach.
    
    \item \textbf{global-EM:} 
    Since the genie-knowledge of the covariance matrix is inaccessible in a practical scenario, we additionally evaluate the \ac{em} algorithm when assuming the prior is Gaussian; thus, in this case, the sample covariance matrix of the training dataset, i.e., $\hat{\B C} = \frac{1}{T_{\text{train}}} \sum_{t=1}^{T_{\text{train}}} \B h_t\B h_t\h$, is used in the M-step \eqref{eq:m_step_sol}.
    
    \item \textbf{genie-BLMMSE:}
    Similar to the \textbf{genie-EM} approach, we assume genie-knowledge of the covariance matrix $\B C_{\B h|\B \delta}$ from \eqref{eq:3gpp_cov} for each pilot observation to evaluate a bound on the performance of the linear \ac{mmse} estimator based on the Bussgang decomposition, cf. \cite{fesl24quant}.
    
    \item \textbf{global-BLMMSE:}
    In this case, instead of utilizing the ground-truth covariance matrix, we utilize the sample covariance matrix for evaluating an approximation of the Bussgang linear \ac{mmse} estimator \cite{fesl24quant}. 
    
    \item \textbf{EM-GM-GAMP:}
    In \cite{Mo2018}, a \ac{cs}-based channel estimator is proposed, which is a combination of an \ac{em} algorithm for approximating the channel \ac{pdf} in the (approximately) sparse angular domain and the \ac{gamp} algorithm to solve the sparse recovery problem. 

    \item \textbf{GMM-BLMMSE:}
    We additionally compare to the parameterized conditional Bussgang estimator based on the \ac{gmm} from \cite{fesl24quant}. Note that learning the channel distribution via a \ac{gmm} is similar to the proposed approach; however, we do not parameterize a linear \ac{mmse} estimator.
    
\end{itemize}

\section{Numerical Results}
For all data-aided approaches, we utilize the same training dataset consisting of $T_{\text{train}} = 100{,}000$ data samples. As performance measure we evaluate an estimate of the normalized \ac{mse} $\E[\|\B h - \hhat\|_2^2] / \E[\|\B h\|_2^2]$ using $10{,}000$ unseen channel samples.
If not otherwise stated, we utilize a \ac{gmm} with $K=64$ full covariances.
We set $L_{\text{max}} = 1{,}000$ and $\gamma=10^{-3}$.

\begin{figure}[t]
	\centering
    \includegraphics[width=0.98\columnwidth]{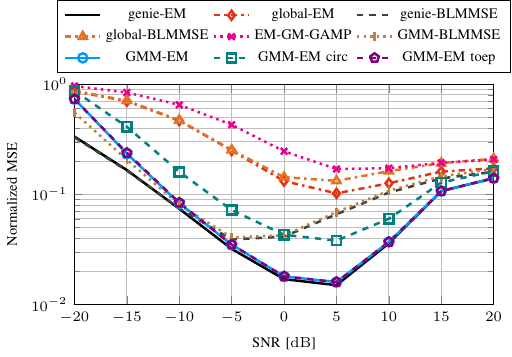}
    \includegraphics[width=0.98\columnwidth]{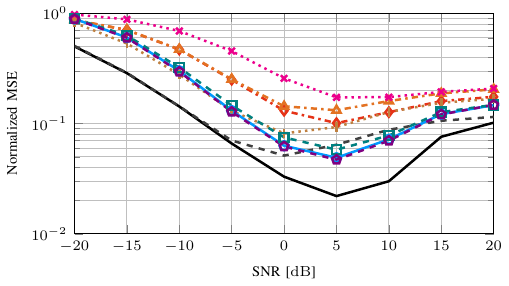}
	\caption{\ac{mse} performance with one (top) and three (bottom) propagation cluster, $N=64$ antennas, and $P=16$ pilot observations.}
     \label{fig:mse}
\end{figure}

Fig. \ref{fig:mse} evaluates the \ac{mse} performance over the \ac{snr} for one (top) and three (bottom) propagation clusters with $N=64$ antennas and $P=16$ pilot observations. It can be observed that \textbf{genie-EM} outperforms \textbf{genie-BLMMSE}, indicating that the \ac{map} objective is superior over the linear \ac{mmse} estimator based on the Bussgang decomposition, especially in the medium to high \ac{snr} regime. This observation is in accordance with the finding in \cite{fesl23cme} that the Bussgang estimator significantly deteriorates from the \ac{cme} in general, justifying the adoption of nonlinear estimation approaches. 
The proposed \textbf{GMM-EM} is on par with the related \textbf{GMM-BLMMSE} in the low \ac{snr} regime; however, in medium and high \ac{snr}, it exhibits a significant performance improvement, even outperforming the \textbf{genie-BLMMSE} approach, which requires genie knowledge of the true covariance matrix for each observation. 
Moreover, all other state-of-the-art estimators are outperformed with a large gap over the whole range of \ac{snr}. In the case of one propagation cluster, the proposed \textbf{GMM-EM} approach even reaches the performance of \textbf{genie-EM}. 
In contrast, the gap is generally higher for more propagation clusters where the channel distribution is less structured. The Toeplitz-structured \ac{gmm} is on par with the full \ac{gmm}, whereas there is a larger gap for the circulant case, imposing a stronger~approximation.

 \begin{figure}[t]
	\centering
    \includegraphics[width=0.98\columnwidth]{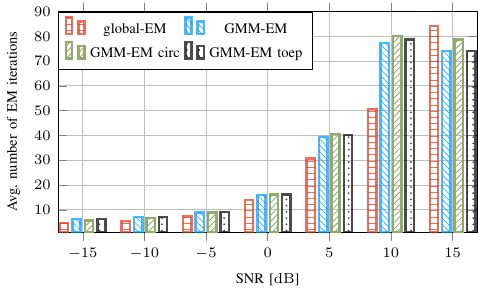}
	\caption{Average number of \ac{em} iterations until convergence for one propagation cluster, $N=64$ antennas, and $P=16$ pilot observations.}
     \label{fig:iterations}
\end{figure}

In Fig. \ref{fig:iterations}, we analyze the necessary number of \ac{em} iterations until convergence for the setup of one propagation cluster in Fig.~\ref{fig:mse}. The number of iterations is particularly low in the low \ac{snr} regime, i.e., below ten iterations, for all considered variants utilizing the \ac{em} algorithm. This is a convenient property since the considered operating range of low-resolution systems is the low \ac{snr} regime where the capacity is not severely reduced \cite{10.1145/1143549.1143827}.

\begin{figure}[t]
	\centering
    \includegraphics[width=0.98\columnwidth]{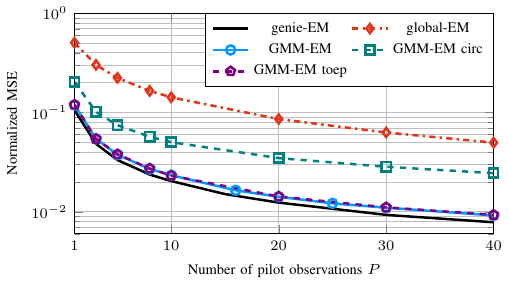}
	\caption{\ac{mse} performance for varying numbers of pilot observations for one propagation cluster, $N=64$ antennas, and $\text{SNR} = \qty{5}{dB}$.}
     \label{fig:pilots}
\end{figure}

Fig. \ref{fig:pilots} investigates the \ac{mse} performance for a varying number of pilot observations for $N=64$ antennas and an \ac{snr} of $\qty{5}{dB}$. 
Since the \textbf{EM-GM-GAMP} is consistently worse than \textbf{global-EM}, and \textbf{GMM-BLMMSE} suffers from a too high complexity in the large pilot regime, we leave out these comparisons.
It can be observed that the proposed \textbf{GMM-EM} approach is almost on par with the genie-aided variant over the whole range. In conclusion, the number of pilots can be drastically reduced when using the proposed approach compared to a simplistic Gaussian prior without performance losses. This, in turn, results in an increased data rate, better energy efficiency, and a lower latency of the channel estimation.

\begin{figure}[t]
	\centering
    \includegraphics[width=0.98\columnwidth]{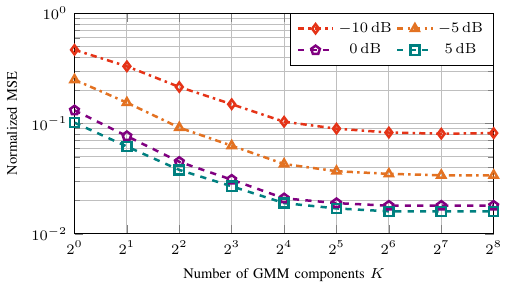}
	\caption{\ac{mse} performance for different numbers of \ac{gmm} components for one propagation cluster, $N=64$ antennas, and $P=16$ pilot observations.}
     \label{fig:components}
\end{figure}

Finally, Fig. \ref{fig:components} evaluates the \ac{mse} performance for varying numbers of \ac{gmm} components of the \textbf{GMM-EM} approach for one propagation cluster, $N=64$ antennas, and $P=16$ pilot observations. The performance drastically improves from $K=1$ (equal to the \textbf{global-EM} variant) to $K=16$. Afterward, a saturation occurs where even more \ac{gmm} components only lead to marginal improvements.

\section{Conclusion and Outlook}

We have presented a framework for significantly enhancing a classical channel estimation algorithm for quantized systems based on a conditionally Gaussian latent generative prior, i.e., a \ac{gmm}. 
Inferring the component with the highest responsibility for a given pilot observation allows for having a strong prior together with a low-complexity implementation due to closed-form analytic solutions of the involved optimization problems of the \ac{em} algorithm. Furthermore, we outlined how structurally constrained \ac{gmm} covariances reduce the memory and complexity overhead.

In future work, we aim to extend the presented framework to higher resolution quantization \cite{mezghani_em} and different generative priors based on conditionally Gaussian latent models, e.g., \acp{vae} and \acp{mfa}. Additionally, we consider learning the generative prior from quantized pilot observations as training data~\cite{fesl24quant}.

\bibliographystyle{IEEEtran}
\bibliography{IEEEabrv,biblio}

\end{document}